\documentclass[conference,twocolumn]{IEEEtran}
\IEEEoverridecommandlockouts
\usepackage{cite}
\usepackage{amsmath,amssymb,amsfonts}
\usepackage{algorithm}
\usepackage{algorithmic}
\usepackage{graphicx}
\usepackage{textcomp}
\usepackage{xcolor}
\usepackage{amsthm}
\usepackage{dsfont}
\usepackage{setspace}
\usepackage[T1]{fontenc}
\usepackage{aecompl}
\usepackage{url}

\usepackage[final]{changes}   

\def\BibTeX{{\rm B\kern-.05em{\sc i\kern-.025em b}\kern-.08em
    T\kern-.1667em\lower.7ex\hbox{E}\kern-.125emX}}
    
\newtheorem{theorem}{Theorem}
\newtheorem{lemma}{Lemma}

\begin{document}

\title{Optimal Update for Energy Harvesting Sensor with Reliable Backup Energy\\
}

\author{
	\IEEEauthorblockN{Lixin Wang$^1$, Fuzhou Peng$^2$, Xiang Chen$^2$, \textit{Member, IEEE}, Shidong Zhou$^1$, \textit{Member, IEEE}}
	\IEEEauthorblockA{$^{1}$Department of Electronic Engineering, Tsinghua University, Beijing, Beijing 100084, China}
	\IEEEauthorblockA{$^{2}$School of Electronics and Information Technology, Sun Yat-sen University, Guangzhou, Guangdong 510006, China}
	\IEEEauthorblockA{wanglx19@mails.tsinghua.edu.cn, pengfzh@mail2.sysu.end.cn, chenxiang@mail.sysu.end.cn, zhousd@tsinghua.end.cn}
}
	

\maketitle

\begin{abstract}
In this paper, we consider an information update system where a wireless sensor sends timely updates to the destination over an erasure channel with the supply of harvested energy and reliable backup energy. The metric Age of Information(AoI) is adopted to measure the timeliness of the received updates at the destination. We aim to find the optimal information updating policy that minimizes the time-average weighted sum of the AoI and the reliable backup energy cost by formulating an infinite state Markov decision process(MDP). The optimal information updating policy is proved to have a threshold structure. Based on this special structure, an algorithm for efficiently computing the optimal policy is proposed. Numerical results show that the optimal updating policy proposed outperforms baseline policies.
\end{abstract}


\section{Introduction}
\label{section1}

Timely information updates from wireless sensors to the destination are critical in real-time monitoring and control systems.
In order to describe the timeliness of information updates, the metric Age of Information(AoI) is proposed\cite{kaul2012real}. Different from general performance metrics such as delay and throughput, AoI refers to the time elapsed since the generation of the latest received information. A lower AoI usually reflects the more timely information received by the destination.    
Therefore, the AoI-minimal status updating policies in sensor networks have been widely studied\cite{sun2017update,kadota2018scheduling,tang2020minimizing}.

In sensor-based information updating systems, energy is consumed in the process of sensing and transmitting updates. If the sensor's energy comes from the grid, it pays the electricity bill. If the sensor's energy comes from its own non-rechargeable battery, the price of sensing and transmitting updates is the cost of frequent battery replacement. We call these sources \textit{reliable energy} since they enable sensors reliable to operate until the power grid is cut off or sensors' batteries are exhausted. There is clearly a price to be paid for using reliable energy to update. 

In order to reduce the reliable energy consumption, a reasonable idea is to introduce energy harvesting technology\cite{ma2019sensing}. Energy harvesting can continuously replenish energy for the sensor by extracting energy from solar power, ambient RF and thermal energy. The harvested energy is stored in the sensor’s rechargeable battery. Since the harvested energy is renewable, it can be used for free. Hence, reliable energy can serve as backup energy. The design of coexistence of reliable backup energy and harvested energy has been researched and promoted in academia and industry\cite{jackson2019capacity,instruments2019bq25505}. However, because the harvested energy arrives sporadically and irregularly, and the capacity of rechargeable batteries is limited, we still need to schedule the usage of energy properly to reduce the cost of using reliable backup energy while maintaining the timeliness of information updates(i.e. the average AoI).

Intuitively, the average AoI and the cost of using reliable energy cannot be minimized simultaneously. On the one hand, a lower average AoI means that the sensor senses and transmits updates more frequently, which will increase the consumption of reliable backup energy since the harvested energy is limited. On the other hand, to reduce the cost of reliable backup energy, the sensor will only exploit the harvested energy. Due to the uncertainty of the energy harvesting behavior, the average AoI of the system will inevitably increase.

Therefore, in this paper, we focus on achieving the best trade-off between the average AoI and the cost of reliable backup energy in a sensor-based information update system where an energy harvesting sensor with reliable backup energy sends timely updates to the destination through an erasure channel. Related work includes\cite{wu2017optimal,bacinoglu2018achieving,arafa2019age,wu2020optimal,wu2020delay,draskovic2021optimal}. \cite{wu2017optimal,bacinoglu2018achieving,arafa2019age} investigate AoI-minimal status updating policies for sensor networks that rely solely on harvested energy. In \cite{wu2020optimal,wu2020delay,draskovic2021optimal}, although the sensors can use both harvested energy and reliable energy, the authors only optimize for delay or throughput and ignore the timeliness of the system. Based on our settings, we will minimize the long-term average weighted sum of the AoI and the paid reliable energy cost to find the optimal information updating policy. The structure of the optimal policy will be analyzed theoretically, and its performance will be demonstrated through simulation.

\section{SYSTEM MODEL and Problem Formulation}
\label{section2}
\subsection{System Model Overview}

\begin{figure}[tbp]
\centerline{\includegraphics[width=0.5\textwidth]{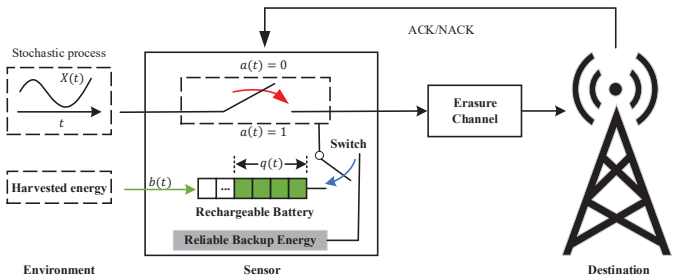}}
\caption{System model.}
\label{system}
\end{figure}

In this paper, we consider a point-to-point information update system where a wireless sensor and a destination are connected by an erasure channel, as shown in Fig.~\ref{system}. Wireless sensors can use the free harvest energy stored in the rechargeable battery and the reliable backup energy that needs to be paid to generate and send real-time environmental status information. The destination keeps track of the environment status through the received updates. We apply the metric Age of Information to measure the freshness of the status information available at the destination.

Without loss of generality, time is slotted with equal length and indexed by $t\in\mathbb N$. 
At the beginning of each time slot, the sensor decides whether to generate and transmit an update to the destination or stay idle. 
The decision action at slot $t$, denoted by $a[t]$, takes value from action set $\mathcal{A}=\left\{0,1\right\}$, where $a[t] =1$ means that the sensor decides to generate and transmit an update to the destination while $a[t]=0$ means the sensor is idle. The channel between the sensor and the destination is assumed to be noisy and time-invariant, and each update will be corrupted with probability $p$ during transmission (Note $p \in (0,1)$). The destination will feed back an instantaneous ACK to the sensor through an error-free channel when it has successfully received an update and a NACK otherwise. We assume the above processes can be completed in one time slot.



\subsection{Age of Information}
Age of Information (AoI) is defined as the elapsed time since the generation of the latest successfully received update in this paper. 
Let $U[t]$ be the time slot when the most recently received update is generated before time slot $t$, and $\Delta[t]$ denote the AoI of destination in time slot $t$. Then, the AoI is given by
\begin{equation}
{\Delta[t]} = t - U[t].
\label{AoI}
\end{equation}

\begin{figure}[tbp]
\centerline{\includegraphics[width=0.5\textwidth]{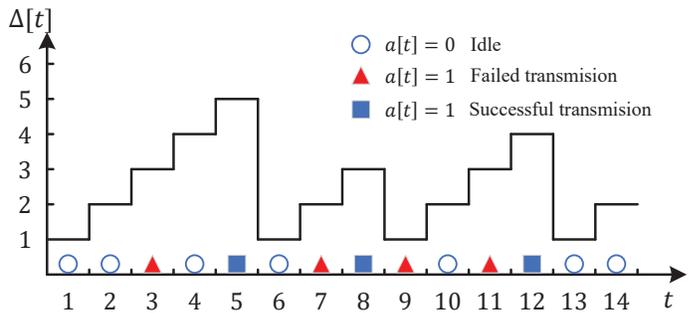}}
\caption{A sample path of AoI with initial age 1.}
\label{AoI_evolution}
\end{figure}

In particular, the AoI will decrease to one if a new update is successfully received. Otherwise it will increase by one. To summarize, the evolution of AoI can be expressed as follows:

\begin{equation}
\label{equ_evolving_AoI}
\Delta[t+1]=
\begin{cases}
1, &\text{ successful transmission},\\
\Delta[t]+1, &\text{ otherwise}.
\end{cases}
\end{equation}
A sample path of AoI is depicted in Fig.~\ref{AoI_evolution}.

\subsection{Description of Energy Supply}
We assume that only the sensor's measurement and transmission process will consume energy, and other energy consumption is ignored.
 The energy unit is normalized, so the generation and transmission for each update will consume one energy unit. As previously described, the energy sources of the sensor include energy harvested from nature and reliable backup energy. The sensor can store the harvested energy in a rechargeable battery for later use. The maximum capacity of the rechargeable battery is $B$ units ($B > 1$). Let $b(t)$ be the accumulated harvested energy in time slot $t$. Since the energy to be harvested is relatively limited, sometimes $b(t)$ does not reach an energy unit. So we consider using the Bernoulli process with the parameter $\lambda$ to approximately capture the arrival process of harvested energy, which is also adopted in \cite{valentini2016optimal,dong2020energy,gindullina2021age}. That is, we have $\Pr\left\{b(t)=1\right\}= \lambda$ and $\Pr\left\{b(t)=0\right\}= 1-\lambda$ in each time slot $t$.
For reliable backup energy, we assume that it contains much more energy units than the rechargeable battery can store, so the energy it contains is infinite. However, it needs to be used for a fee. Therefore, when the power of the rechargeable battery is not 0, the sensor will prioritize using the energy in the rechargeable battery for status update, otherwise, it will automatically switch to the reliable backup energy until the sensor has harvested energy. Defining the power of the rechargeable battery at the beginning of time slot $t$ as the battery state $q[t]$, then the evolution of battery state between time slot $t$ and $t+1$ can be summarized as follows:
\begin{equation}
\label{q_evolution}
q[t+1] = \min \{ q[t] + b[t] - a[t]u(q[t]),B\} ,
\end{equation}
where $u(\cdot)$ is unit step function, which is defined as
\begin{equation}
\label{indicator}
u(x)=
\begin{cases}
1,&\text{if $x>0$},\\
0,&\text{otherwise}.
\end{cases}
\end{equation}
Suppose that under paid reliable energy supply, the cost of generating and transmitting an update is a non-negative value $C_r$. Define $E[t]$ as the paid reliable energy costs at the time slot $t$, then we have
\begin{equation}
\label{E_cost}
E[t] = {C_r}a[t](1-u(q[t])).
\end{equation}




\subsection{Problem Formulation}
Let $\Pi$ denote the set of non-anticipated policies in which scheduling decision $a[t]$ are made based on the action history $\left\{a[k]\right\}_{k=0}^{t-1}$, the AoI evolution $\left\{\Delta[k]\right\}_{k=0}^{t-1}$, the evolution of battery state $\left\{q[k]\right\}_{k=0}^{t-1}$  as well as the system parameters(i.e. $p$, $\lambda$, etc.).
In order to keep the information freshness at the destination, the sensor needs to send updates. 
However, due to the randomness of harvested energy arrivals, the battery energy may sometimes be insufficient to support updates, and the sensor has to take energy from reliable backup energy. To balance the information freshness and the paid reliable backup energy costs, we aim to find the optimal information updating policy $\pi \in \Pi$ that achieves the minimum of the time-average weighted sum of the AoI and the paid reliable backup energy costs. 
The problem is formulated as follows:
\begin{equation}
\begin{aligned}
\label{problem}
&\mathop {\min }\limits_{\pi  \in \Pi } \mathop {\lim\sup }\limits_{T \to \infty }  \frac{1}{T}{\mathbb E}\left\{\sum\limits_{t = 0}^{T-1}  [{\Delta [t]} + \omega E[t]]\right\}, \\
&\text{s}.\text{t}.\qquad (2),(3),(5), \\
\end{aligned}
\end{equation}
where $\omega$ is the positive weighting factor.



\section{Optimal policy analysis}
\label{section3}
In this section, we aim to solve the problem (\ref{problem}) and obtain the optimal policy. It is difficult to solve the original problem directly due to the random erasures and the temporal dependency in both AoI evolution and battery state evolution. So we reformulate the original problem as a time-average cost MDP with infinite state space and analyze the structure of the optimal policy.

\subsection{Markov Decision Process Formulation}
According to the system description mentioned in the previous section, the MDP is formulated as follows:

\begin{itemize}
\item \textbf{State Space}. The state of a sensor $\textbf{x}[t]$ in slot $t$ is a couple of the current destination-AoI and the battery state, i.e., $(\Delta[t], q[t])$. Define $\mathcal{B}=\left\{0,1,...,B \right\}$. The state space $\mathcal{S}= \mathbb Z^+ \times \mathcal{B}$ is thus inﬁnite countable.
\item \textbf{Action Space}. The sensor's action $a[t]$ in time slot $t$ only takes value from the action set $\mathcal{A}=\left\{0,1\right\}$. 
\item \textbf{Transition Probabilities}. Denote $\Pr ({\textbf{x}}[t + 1]|{\textbf{x}}[t],a[t])$ as the transition probability that current state $\textbf{x}[t]$ transits to next state $\textbf{x}[t+1]$ after taking action $a[t]$. Suppose the current state ${\textbf{x}}[t] = (\Delta, q)$ and action $a[t] = a$, then the transition probability is divided into two following cases conditioned on different values of action.

\textbf{\emph{Case 1}}. $a=0$,
\begin{equation}
\label{transition_case1_v2}
\begin{cases}
\Pr \{ (\Delta+1, q+1)|(\Delta, q),0\}=\lambda, &\text{ if }q < B, \\
\Pr \{(\Delta+1, B)|(\Delta, B),0\}= 1,         &\text{ if } q = B, \\
\Pr \{(\Delta+1, q)|(\Delta, q),0\}= 1-\lambda, &\text{ if }q < B. \\
\end{cases}
\end{equation}

\textbf{\emph{Case 2}}. $a=1$,
\begin{equation}
\label{transition_case2_v2}
\begin{cases}
\Pr \{ (\Delta+1, q)|(\Delta, q),1\}=p\lambda,              &\text{ if }q>0, \\
\Pr \{ (1, q)|(\Delta, q),1\}=(1-p)\lambda,                 &\text{ if }q>0, \\
\Pr \{\Delta+1, q-1)|(\Delta, q),1\}=p(1-\lambda),           &\text{ if }q>0, \\
\Pr \{ (1, q-1)|(\Delta, q),1\}=(1-p)(1-\lambda),           &\text{ if }q >0, \\
\Pr \{ (\Delta+1, 1)|(\Delta, 0),1\}=p\lambda,              &\text{ if }q=0, \\
\Pr \{ (1, 1)|(\Delta, 0),1\}=(1-p)\lambda,                 &\text{ if }q=0, \\
\Pr \{(\Delta+1, 0)|(\Delta, 0),0\}= p(1-\lambda),          &\text{ if } q = 0, \\
\Pr \{(1, 0)|(\Delta, 0),0\}= (1-p)(1-\lambda),             &\text{ if } q = 0. \\
\end{cases}
\end{equation}
In both cases, the evolution of AoI still follows equation \eqref{equ_evolving_AoI} and the evolution of battery state follows \eqref{q_evolution}.

\item \textbf{One-step Cost}. For the current state $\textbf{x}=(\Delta, q)$, the one-step cost $C(\textbf{x},a)$ of taking action $a$ is expressed by
 
\begin{equation}    
    \label{onestepcost}
    C(\textbf{x},a) = \Delta  + \omega {C_r}a(1-u(q)).
\end{equation}
\end{itemize}

After the above modeling, the original problem \eqref{problem} is transformed into obtaining the optimal policy for the MDP to minimize the average cost in an infinite horizon:
\begin{equation}
\label{trans_problem}
 \mathop {\lim\sup }\limits_{T \to \infty }  \frac{1}{T}{\mathbb E_\pi}\left\{ \sum\limits_{t = 0}^{T-1} C(\textbf{x}[t],a[t])\right\}.
\end{equation}
Denote $\Pi_{SD}$ as the set of stationary deterministic policies. Given observation$(\Delta[t],q[t])=(\Delta,q)$, the policy $\pi \in \Pi_{SD}$ selects action $a[t]=\pi(\Delta,q)$, where $\pi(\cdot):(\Delta,q)\to\left\{0,1\right\}$ is a deterministic function from state space $\mathcal{S}$ to action space $\mathcal{A}$.
According to \cite{altman1999constrained} , there exists a stationary deterministic policy to minimize the above unconstrained MDP with inﬁnite countable state and action space under certain verifiable conditions. In the next section, the structural properties of the optimal policy are investigated.
\subsection{Structure Analysis of Optimal Policy}

According to \cite{sennott1989average}, there exits a value function $V(\textbf{x})$ which satisﬁes the following Bellman equation for the infinite horizon average cost MDP:
\begin{equation}
\lambda + V(\textbf{x}) = \min_{a \in \mathcal{A}} \left\{ C(\textbf{x},a) + \sum_{\textbf{x}^\prime \in \mathcal{S}} \Pr (\textbf{x}^\prime|\textbf{x},a)V(\textbf{x}^\prime) \right\},
\label{bel_equation}
\end{equation}
where $\lambda$ is the average cost by following the optimal policy. Denote $Q(\textbf{x},a)$ as the state-action value function which means the value of taking action $a$ in state $\textbf{x}$. We have:
\begin{equation}
    Q(\textbf{x},a)=C(\textbf{x},a) + \sum_{\textbf{x}^\prime \in \mathcal{S}} \Pr (\textbf{x}^\prime|\textbf{x},a)V(\textbf{x}^\prime).
\label{Qfunction}
\end{equation}

So the optimal policy $\pi ^\star \in \Pi_{SD}$ in state $\textbf{x}$ can be expressed as follows:
\begin{equation}
    {\pi ^\star}(\textbf{x}) = \arg \mathop {\min }\limits_{a \in \mathcal{A}} Q(\textbf{x},a).
\label{piandQequation}
\end{equation}

Next, we first prove the monotonicity of the value function on different dimensions, which is summarized in the following lemma.

\begin{lemma}
    For a fixed channel erasure probability $p$, given the battery state $q$ and for any $1\leq \Delta_1\leq\Delta_2$, we have 
    \begin{equation}
    \label{lemma1_part1}
        V(\Delta_1,q)\le V(\Delta_2,q),
    \end{equation}
    and, given AoI $\Delta\geq 1$,
    \begin{equation}
        \label{lemma1_part2}
        V(\Delta,q)\ge V(\Delta,q+1)
    \end{equation}
    holds for any $q\in \left\{0,1,...,B-1\right\}$.
    \label{lemma1}
\end{lemma}

\begin{IEEEproof}
See Appendix \ref{app_proof_lemma_monitonic} in Supplementary Material \cite{tsinghua.edu}.
\end{IEEEproof}

Based on Lemma \ref{lemma1}, we then establish the incremental property of the value function, which is shown in the following lemma.

\begin{lemma}
    For a fixed channel erasure probability $p$, for any $\Delta_1 \le \Delta_2$ and given $q \in \mathcal{B}$, we have:
    \begin{equation}
    \label{lemm2_formula1}
        V(\Delta_2,q)-V(\Delta_1,q)\ge \Delta_2-\Delta_1.
    \end{equation}
    And, for any $q \in \left\{0,1,...,B-1\right\}$ and $\Delta \in \mathbb Z^+ $, we have:
    \begin{equation}
    \label{lemm2_formula2}
        V(\Delta+1,q+1)-V(\Delta,q+1)\ge p[V(\Delta+1,q)-V(\Delta,q)].
    \end{equation}
    \label{lemma2}    
\end{lemma}
\begin{IEEEproof}
See Appendix \ref{app_proof_lemma_creasement} in Supplementary Material \cite{tsinghua.edu}. 
\end{IEEEproof}

With Lemma \ref{lemma1} and Lemma \ref{lemma2}, we directly provide our main result in the following Theorem.

     
\begin{theorem}
    Assuming that the channel erasure probability $p$ is fixed. For given battery state $q$, there exists a threshold $\Delta_q$ , such that when $\Delta\ < \Delta_q$, the optimal action $\pi^\star (\Delta,q)=0$, i.e., the sensor keeps idle; when $\Delta \ge \Delta_q$, the optimal action $\pi^\star(\Delta,q)=1$, i.e., the sensor chooses to generate and transmit a new update.
\label{theorem1}
\end{theorem}
\begin{IEEEproof}
The optimal policy is of a threshold structure if $Q(\textbf{x},a)$ has a sub-modular structure, that is,
\begin{equation}
Q(\Delta,q,0)- Q(\Delta,q,1) \leq  Q(\Delta+1,q,0)- Q(\Delta+1,q,1).
\end{equation}

We will divide the whole proof into the following three cases:
    
    
    \textbf{Case 1}. When $q=0$, for any $\Delta \in \mathbb Z^+$ we have:
    \begin{align}
        &Q(\Delta,q,0)-Q(\Delta,q,1)\nonumber\\
        =&\Delta+ \lambda V(\Delta+1,q+1)+(1-\lambda)V(\Delta+1,q)\nonumber\\
        &-\Delta-\omega{C_r}-p\lambda V(\Delta+1,q+1)+p(1- \lambda)V(\Delta+1,q)\nonumber\\
    &-(1-p)\lambda V(1,q+1)-(1-p)(1-\lambda) V(1,q)\nonumber\\
    =&(1-p)\lambda(V(\Delta+1,q+1)-V(1,q+1))\nonumber\\
    &+(1-p)(1-\lambda)(V(\Delta+1,q)-V(1,q))-\omega{C_r}.
    \end{align}
    
Therefore, we have
\begin{align}
&Q(\Delta+1,q,0)-Q(\Delta+1,q,1) - [Q(\Delta,q,0)-Q(\Delta,q,1)]\nonumber\\
=&(1-p)\lambda(V(\Delta+2,q+1)-V(\Delta+1,q+1)) \nonumber\\
&+(1-p)(1-\lambda)(V(\Delta+2,q)-V(\Delta,q))\nonumber\\
\overset{(a)}{\geq}& 0,
\end{align}
where the last inequality $(a)$ is due to the monotonicity property revealed by \eqref{lemma1_part1} in Lemma \ref{lemma1}. 

    \textbf{Case 2}. When $q \in \left\{1,...,B-1\right\}$,for any $\Delta \in \mathbb Z^+$ we have:
    \begin{align}
        &Q(\Delta+1,q,0)-Q(\Delta+1,q,1)-[Q(\Delta,q,0)-Q(\Delta,q,1)]\nonumber\\
        =&Q(\Delta+1,q,0)-Q(\Delta,q,0)-[Q(\Delta+1,q,1)-Q(\Delta,q,1)]\nonumber\\
        =&\lambda[V(\Delta+2,q+1)-V(\Delta+1,q+1)]\nonumber\\
        &-p\lambda[V(\Delta+2,q)-V(\Delta+1,q)]\nonumber\\
        &+(1-\lambda)[V(\Delta+2,q)-V(\Delta+1,q)]\nonumber\\
        &-p(1-\lambda)[V(\Delta+2,q-1)-V(\Delta+1,q-1)]\nonumber\\
        \overset{(a)}{\geq}& 0,
        \label{submodular}
    \end{align}    
    where the last inequality $(a)$ is due to the incremental property revealed by \eqref{lemm2_formula2} in Lemma \ref{lemma2}. 
    
    
    \textbf{Case 3}. When $q=B$,for any $\Delta \in \mathbb Z^+$ we have:
    \begin{align}
        &Q(\Delta+1,q,0)-Q(\Delta+1,q,1)-[Q(\Delta,q,0)-Q(\Delta,q,1)]\nonumber\\
        =&Q(\Delta+1,q,0)-Q(\Delta,q,0)-[Q(\Delta+1,q,1)-Q(\Delta,q,1)]\nonumber\\
        =&(1-\lambda)[V(\Delta+2,q)-V(\Delta+1,q)]\nonumber\\
        &-p(1-\lambda)[V(\Delta+2,q-1)-V(\Delta+1,q-1)]\nonumber\\
        \overset{(a)}{\geq}& 0,
        \label{B_submodular}
    \end{align}     
    where the last inequality (a) is also due to the incremental property revealed by \eqref{lemm2_formula2} in Lemma \ref{lemma2}.
    
    
Therefore, we have completed the whole proof.
\end{IEEEproof}
Theorem \ref{theorem1} reveals the threshold structure of the optimal policy: if the optimal action in a certain state is to generate and transmit an update, then in the state with the same battery state and larger AoI, the optimal action must be the same. 

Based on this unique threshold structure, we propose a modified value iteration algorithm to solve the optimal policy, as shown in Algorithm \ref{mvia}. Specifically, We first iterate the Bellman equation \eqref{bel_equation} to obtain the value function. Then based on the threshold structure, \added{the optimal policy can be obtained without calculating the equation \eqref{piandQequation} in each state}\deleted{we can find the optimal policy by equation \eqref{piandQequation} without traversing all the states}, which reduces the computational complexity.

\begin{algorithm}[tb]
    \caption{Modified Value Iteration Algorithm}
    \label{mvia}
    \begin{algorithmic}[1]
        \REQUIRE ~~\\ 
        Iteration number $K$ and iteration threshold $\epsilon$.
        \ENSURE ~~\\ 
        Optimal policy $\pi^\star(\textbf{x})$ for all state $\textbf{x}$.
        \STATE \textbf{Initialization: }$V_0(\textbf{x})= 0.$
        \FOR{episodes $k = 0,1,2,...,K$}
            \FOR{state $\textbf{x}\in \mathcal{S}$}
                \FOR{action $a\in \mathcal{A}$}
                    \STATE $Q_k(\textbf{x},a)\leftarrow C(\textbf{x},a) +\underset{\textbf{x}^\prime \in \mathcal{S}}{\sum}   \Pr (\textbf{x}^\prime|\textbf{x},a)V_k(\textbf{x}^\prime)$ 
                \ENDFOR
            \STATE ${V_{k + 1}}(\textbf{x}) \leftarrow  \mathop {\min }\limits_{a \in \mathcal{A}} Q_k(\textbf{x},a)$               
            \ENDFOR
            \IF{$\|V_{k+1}(\textbf{x})-V_{k}(\textbf{x})\|\leq \epsilon$}
                \FOR{$\textbf{x}=(\Delta,q) \in \mathcal{S}$}
                    \IF{$\pi^\star(\Delta-1,q)=1$} 
                    \STATE $\pi^\star(\textbf{x})\leftarrow 1$,
                    \ELSE
                    \STATE ${\pi^\star}(\textbf{x}) \leftarrow \arg \mathop {\min }\limits_{a \in \mathcal{A}} Q_k(\textbf{x},a)$           
                    \ENDIF
                \ENDFOR
            \ENDIF
        \ENDFOR
    \end{algorithmic}
\end{algorithm}

\section{Numerical Result}
\label{section4}
In this section, we first show the threshold structure of optimal policy by the simulation results. Then we compare the performance of the optimal policy with the zero-wait policy, the periodic policy, the randomized policy, the energy first policy under different system parameters such as weighting factor $\omega$, energy harvesting probability $\lambda$ and erasure probability $p$. Note that the zero-wait policy means the sensor generates and transmits an update in every time slot\cite{sun2017update}, while the periodic policy means the sensor periodically generates and sends updates to the destination. The randomized policy refers to that the sensor chooses to send an update or stay idle in each time slot with the same probability. The energy first policy means that the sensor only uses the harvested energy, that is, as long as the battery state is not 0, it will choose to sense and send updates, otherwise it will remain idle. Obviously, the energy first policy will not incur the cost of reliable energy. In our simulation, we assume that the cost of reliable energy $C_r$ for one update equals to $2$ and the maximum battery capacity $B$ equals to $20$. 



\begin{figure}[tbp]
    \centerline{\includegraphics[width=0.5\textwidth]{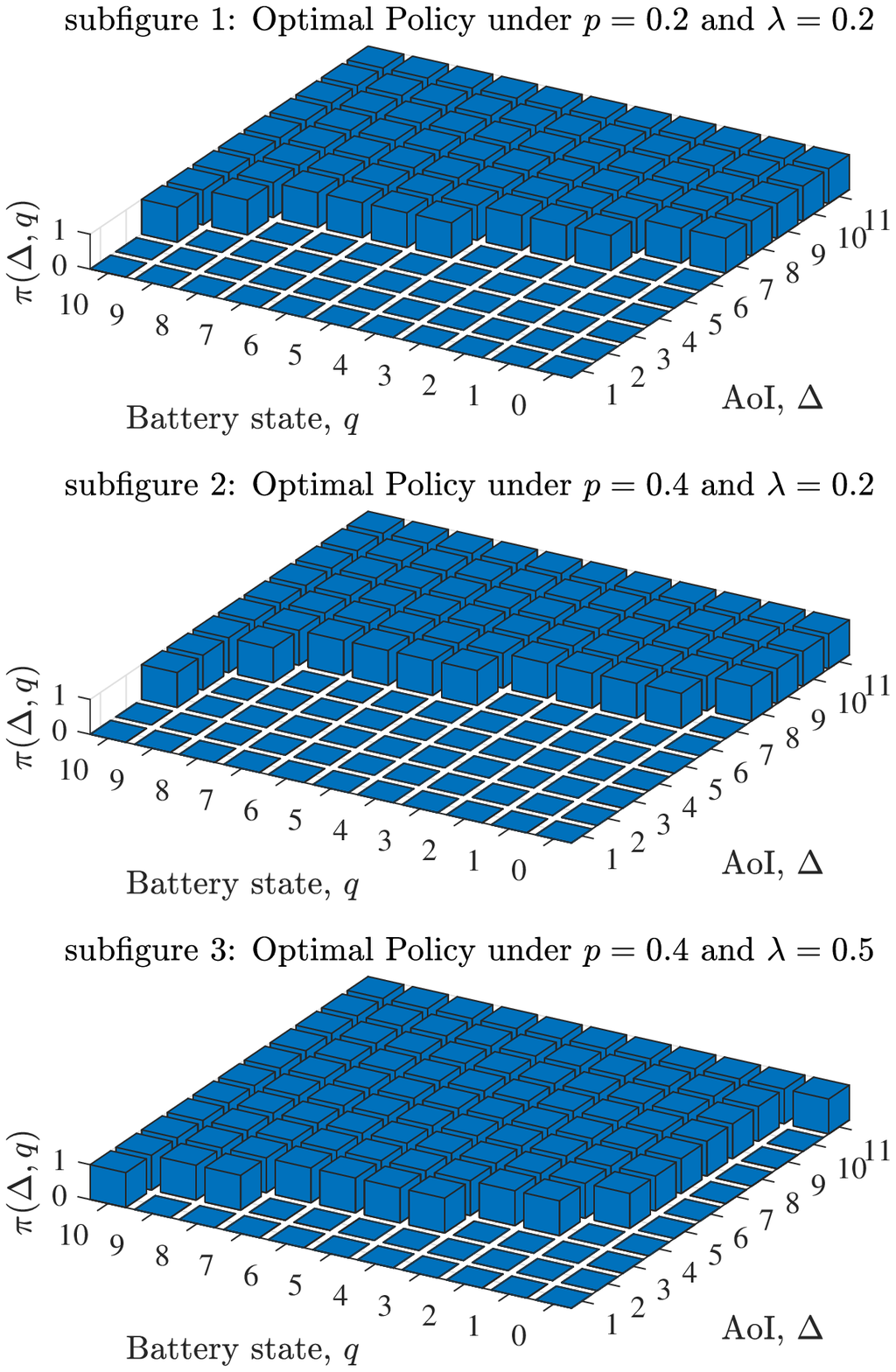}}
    \caption{Optimal policy conditioned on different parameters.}
    \label{fig:thresold_blue}
\end{figure}

Fig.~\ref{fig:thresold_blue} shows the optimal policy under different \replaced{system parameters}{channel erasure probability and energy harvesting probability}.  All the subfigures in Fig.~\ref{fig:thresold_blue}  exhibits the threshold structure described in Theorem \ref{theorem1}. Note that the weighting factor $\omega$ is set to be $10$ , which is neither too small nor too large. Intuitively, when $\omega$ is too small, the optimal action for every state should be 0, and when $\omega$ is too large, the optimal action for every state should be 1.
Fig.~\ref{fig:thresold_blue} shows that when the AoI is small, even if the battery state is not 0, the optimal action in the corresponding state is to keep idle. When the AoI is large or the battery state is large, the optimal action is to measure and send updates.

\begin{figure}[tbp]
\centerline{\includegraphics[width=0.4\textwidth]{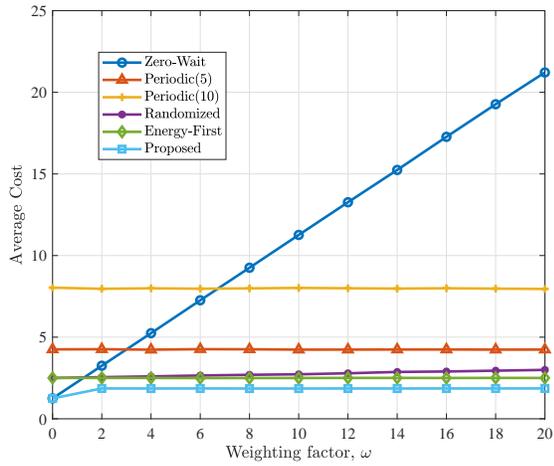}}
\caption{Performance comparison of the zero-wait policy, periodic policy (period = 5), periodic policy (period = 10), randomized policy, energy first policy and proposed policy versus the weighting factor $\omega$ with simulation conditions $p=0.2$, $\lambda=0.5$ and $B=20$.}
\label{costwithb_big}
\end{figure}

\deleted{Then, we show the average cost performance of optimal policy in Fig.~\ref{costwithb_big} under different weighting factor $\omega$.} \deleted{Fig.~\ref{costwithb_big} shows the system performance of the optimal policy under different setting. }
\added{Fig.~\ref{costwithb_big} shows the time average cost under different policies, i.e., the zero-wait policy, the periodic policy, the randomized policy, the energy first policy and the proposed optimal policy.}
\deleted{The optimal policy is compared with the zero-wait policy and the periodic policy \replaced{conditioned on the }{under the same} channel erasure probability $p=0.2$ and the energy harvesting probability $\lambda=0.5$}\deleted{ in this simulation}
Here we set the period \added{of the periodic policy }to 5 and 10 for comparison \added{without loss of generality}. It can be found that under different weighting factor $\omega$, the optimal policy proposed in this paper can obtain the minimum long-term average cost compared with the other policies, which indicates the best trade-off between the average AoI and the cost of reliable energy. When $\omega$ tends to $0$, the zero-wait policy tends to be optimal. \replaced{Since}{When} there is no need to consider the update cost brought by paid reliable backup energy, the optimal policy \replaced{should maximize the utilization of the updating opportunities.}{is to update information in every time slot.}

\begin{figure}[tbp]
\centerline{\includegraphics[width=0.4\textwidth]{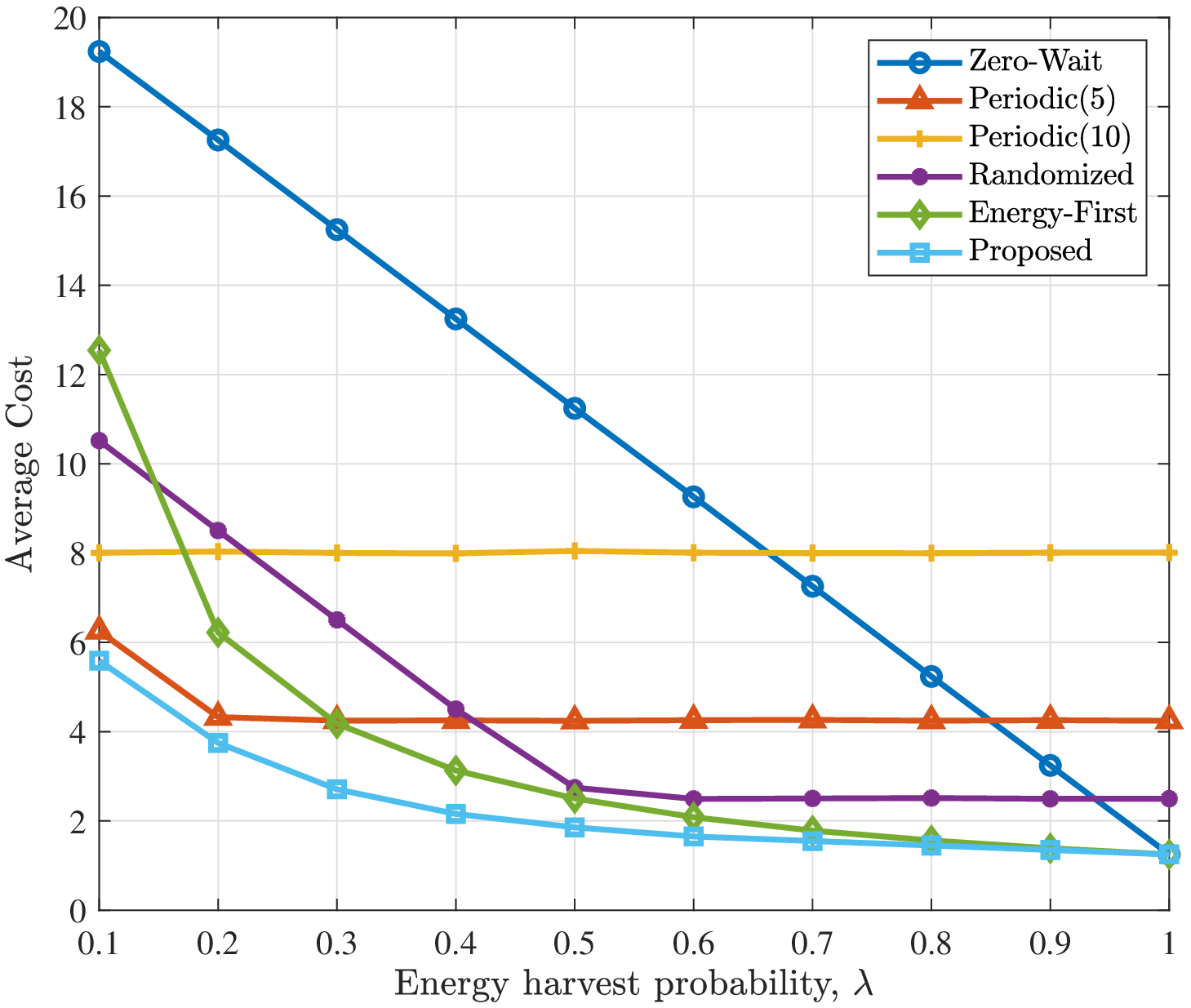}}
\caption{Comparison of the zero-wait policy, periodic policy (period = 5), periodic policy (period = 10), randomized policy, energy first policy and proposed policy versus the energy harvesting probability with simulation conditions $p=0.2$, $\omega=10$ and $B=20$.}
\label{costwithlambda_big}
\end{figure}

\deleted{In }Fig.~\ref{costwithlambda_big}\replaced{ reveals}{, we present} the impact of \deleted{different }energy harvesting probabilities $\lambda$\deleted{ on different policies}. \deleted{In this simulation, we also set }The channel erasure probability $p$ is set to be $0.2$ and weighting factor $\omega$ is $10$\deleted{in this simulation}. It \added{also} can be found \deleted{from Fig.~\ref{costwithlambda_big} }that the proposed optimal update policy outperforms all other policies under different energy harvesting probabilities\deleted{the zero-wait policy and the periodic policy (period = 5)}. The interesting point is that when the probability of energy harvesting tends to 1, \replaced{i.e.}{that is}, energy arrives in each time slot, the performance of the zero-wait policy and the energy first policy is \replaced{equal}{close} to the optimal policy, while there is still a performance gap between the optimal policy and the other two polices. This is intuitive because when the free harvested energy is sufficient, the optimal policy must be to generate and transmit updates in every time slot. However, the periodic policy and the randomized policy still keep idle in many time slots\deleted{send updates in many time slots}, which will \added{lead to a higher average AoI and thus increase the average cost}\deleted{increase the average cost of AoI term and finally increases the whole system cost}\deleted{leads to an increase in the average cost of the system}.

\begin{figure}[tbp]
\centerline{\includegraphics[width=0.4\textwidth]{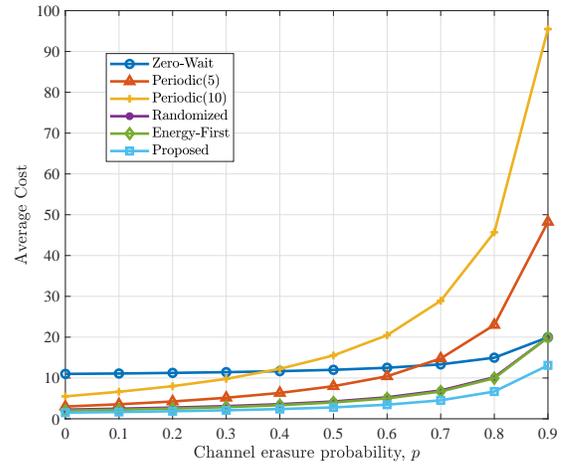}}
\caption{Comparison of the zero-wait policy, periodic policy (period = 5), periodic policy (period = 10), randomized policy, energy first policy and proposed policy versus the erasure probability with simulation conditions $\lambda=0.5$, $\omega=10$ and $B=20$.}
\label{costwithp_big}
\end{figure}

In Fig.~\ref{costwithp_big}, we compare the above five policies under different erasure probabilities $p$. \replaced{The simulation settings are }{In this simulation, we set }the energy harvesting probability $\lambda=0.5$ and weighting factor $\omega=10$. It can be found \deleted{from Fig.~\ref{costwithp_big} }that when erasure probability increases from 0 to 0.9, the proposed optimal update policy always performs better than the other baseline policies. Note when $p = 1$, all the updates are erased by the noisy channel. So it is meaningless to discuss this case.


\section{Conclusion}
\label{section5}
In this paper, we have studied the optimal updating policy for an information update system where a wireless sensor sends updates over an erasure channel using both harvested energy and reliable backup energy. Theoretical analysis indicates the threshold structure of the optimal policy and simulation results verify its performance.

\bibliographystyle{./IEEEtran}
\bibliography{./IEEEexample}


\begin{figure*}
\huge
\begin{center}
Supplementary Material for the paper "Optimal Update for \\Energy Harvesting Sensor with Reliable Backup Energy"    
\end{center}
\end{figure*}

\newpage

\section{APPENDIX}
\subsection{Proof of Lemma \ref{lemma1}}
\label{app_proof_lemma_monitonic}
    The proof requires the use of value iteration algorithm(VIA) and mathematical induction. First, give a brief introduction to VIA, which obtains the value of the value function in different states through continuous iteration. The specific iteration process is as follows:
\begin{equation}
\label{VIA}
\begin{cases}
{{V_0}(\textbf{x}) = {m_\textbf{x}}},\\
    Q_k(\textbf{x},a)=C(\textbf{x},a) +\underset{\textbf{x}^\prime \in \mathcal{S}}{\sum}   \Pr (\textbf{x}^\prime|\textbf{x},a)V_k(\textbf{x}^\prime),\\
{{V_{k + 1}}(\textbf{x}) = \mathop {\min }\limits_{a \in \mathcal{A}} Q_k(\textbf{x},a)},
\end{cases}
\end{equation}
where $m_\textbf{x}$ is an arbitrary initial value of $V_0(\textbf{x})$ with respect to state $\textbf{x}$ and $k \in \mathbb Z$. It’s worth noting that $V_{k + 1}(\textbf{x})$ will converge when $k$ goes into infinity for any state $\textbf{x}$, which can be expressed as follows:
\begin{equation}
\label{Vkinfintiy}
\mathop {\lim }\limits_{k \to \infty } {V_k}(\textbf{x}) = V(\textbf{x}),\forall \textbf{x} \in \mathcal{S}.
\end{equation}

Then we will use mathematical induction to prove the monotonicity of the value function in each component.  

First prove \eqref{lemma1_part1}. At the beginning of the induction method, We need to verify that the inequality $V_1(\Delta_1,q)\le V_1(\Delta_2,q)$  holds when $k=1$. By assuming ${{V_0}(\textbf{x}) = 0},\forall \textbf{x} \in \mathcal{S}$, we have:
\begin{align}
\label{V_0_DELTA1_lemma1}
    V_1(\Delta_1,q) &= \min_{a \in \mathcal{A}}\left\{ Q_0(\Delta_1,q,a)\right\} \nonumber\\
    &=\min\left\{ Q_0(\Delta_1,q,0),Q_0(\Delta_1,q,1)\right\} \nonumber\\
    &=\min\left\{ \Delta_1+\mathbb \omega{C_r} (1-u(q)),\Delta_1\right\} \nonumber\\
    &=\Delta_1,
\end{align}
and,
\begin{align}
\label{V_0_DELTA1_lemma2}
    V_1(\Delta_2,q) &= \min_{a \in \mathcal{A}}\left\{ Q_0(\Delta_2,q,a)\right\} \nonumber\\
    &=\min\left\{ Q_0(\Delta_2,q,0),Q_0(\Delta_2,q,1)\right\} \nonumber\\
    &=\min\left\{ \Delta_2+\omega{C_r} (1-u(q)),\Delta_2\right\} \nonumber\\
    &=\Delta_2.
\end{align}

Therefore, if $\Delta_1\le\Delta_2$, $V_1(\Delta_1,q)=\Delta_1\le\Delta_2=V_1(\Delta_2,q)$.
Then we assume that at the $k$th step of the induction method, the following formula holds:
\begin{equation}
\label{klimit}
    V_k(\Delta_1,q)\le V_k(\Delta_2,q),\forall \Delta_1\le\Delta_2.
\end{equation}
So the next formula that needs to be verified is
\begin{equation}
\label{V_k+1}
    V_{k+1}(\Delta_1,q)\le V_{k+1}(\Delta_2,q),\forall \Delta_1\le\Delta_2
\end{equation}
Since ${{V_{k + 1}}(\textbf{x}) = \mathop {\min }\limits_{a \in \mathcal{A}} Q_k(\textbf{x},a)}$, we need to bring out $Q_k(\textbf{x},a)$ first. The state-action value function $Q_k(\textbf{x},a)$ at state $\textbf{x}=(\Delta, q)$ is as follows:
\begin{equation}
\label{Q_k_equation}
\begin{cases}
 Q_k(\Delta, q,0)=C(\Delta, q,0) +\underset{\textbf{x}^\prime \in \mathcal{S}}{\sum}   \Pr (\textbf{x}^\prime|\textbf{x},0)V_k(\textbf{x}^\prime),\\
    Q_k(\Delta, q,1)=C(\Delta, q,1) +\underset{\textbf{x}^\prime \in \mathcal{S}}{\sum}   \Pr (\textbf{x}^\prime|\textbf{x},1)V_k(\textbf{x}^\prime).
\end{cases}
\end{equation}
Due to the complexity of the transition probability situation and one-step cost function, we will discuss the following three cases:

\textbf{\emph{Case 1}}. $q=0$,

In this case, according to transition probability \eqref{transition_case1_v2} and \eqref{transition_case2_v2}, we have the state-value function $Q_k(\Delta, q,0)$ and $Q_k(\Delta, q,1)$ as follows:
\begin{align}
    Q_k(\Delta, q,0)=\Delta &+\lambda V_k(\Delta+1,q+1) \nonumber\\
    &+(1- \lambda)V_k(\Delta+1,q),
\end{align}
and,
\begin{align}
    Q_k(\Delta, q,1)=\Delta+\omega C_r &+p\lambda V_k(\Delta+1,q+1) \nonumber\\
    &+p(1- \lambda)V_k(\Delta+1,q)\nonumber\\
    &+(1-p)\lambda V_k(1,q+1)\nonumber\\
    &+(1-p)(1-\lambda) V_k(1,q).
\end{align}
Due to that $V_{k}(\Delta,q)$ is assumed to be non-decreasing function with respect to $\Delta$ for any fixed $q$, it is obviously that both $Q_k(\Delta, q,0)$ and $Q_k(\Delta, q,1)$ are non-decreasing with respect to $\Delta$. Therefore,for any $\Delta_1\le\Delta_2$ we have:
\begin{align}
\label{V_K+1_DELTA1_lemma1}
    V_{k+1}(\Delta_1,q) &= \min_{a \in \mathcal{A}}\left\{ Q_k(\Delta_1,q,a)\right\} \nonumber\\
    &=\min\left\{ Q_k(\Delta_1,q,0),Q_k(\Delta_1,q,1)\right\} \nonumber\\
    &\le \min\left\{ Q_k(\Delta_2,q,0),Q_k(\Delta_2,q,1)\right\} \nonumber\\
    &=V_{k+1}(\Delta_2,q).
\end{align}
As a result, with the induction we prove that $V_{k}(\Delta,q)$ is non-decreasing function for any $k$ with respect to $\Delta$ and $q=0$. By taking the limits on both side of \eqref{klimit} we prove that \eqref{lemma1_part1} holds in the case $q=0$.

\textbf{\emph{Case 2}}. $0<q<B$,

In this case, according to transition probability \eqref{transition_case1_v2} and \eqref{transition_case2_v2}, we have the state-value function $Q_k(\Delta, q,0)$ and $Q_k(\Delta, q,1)$ as follows:
\begin{align}
    Q_k(\Delta, q,0)=\Delta &+\lambda V_k(\Delta+1,q+1) \nonumber\\
    &+(1- \lambda)V_k(\Delta+1,q),
\end{align}
and,
\begin{align}
    Q_k(\Delta, q,1)=\Delta &+p\lambda V_k(\Delta+1,q) \nonumber\\
    &+p(1- \lambda)V_k(\Delta+1,q-1)\nonumber\\
    &+(1-p)\lambda V_k(1,q)\nonumber\\
    &+(1-p)(1-\lambda) V_k(1,q-1).
\end{align}
Due to $V_{k}(\Delta,q)$ is assumed to be non-decreasing function with respect to $\Delta$ for any fixed $q$, it is obviously that both $Q_k(\Delta, q,0)$ and $Q_k(\Delta, q,1)$ are non-decreasing with respect to $\Delta$. Therefore,for any $\Delta_1\le\Delta_2$ we have:
\begin{align}
\label{V_K+1_DELTA1_lemma1_case2}
    V_{k+1}(\Delta_1,q) &= \min_{a \in \mathcal{A}}\left\{ Q_k(\Delta_1,q,a)\right\} \nonumber\\
    &=\min\left\{ Q_k(\Delta_1,q,0),Q_k(\Delta_1,q,1)\right\} \nonumber\\
    &\le \min\left\{ Q_k(\Delta_2,q,0),Q_k(\Delta_2,q,1)\right\} \nonumber\\
    &=V_{k+1}(\Delta_2,q).
\end{align}
As a result, with the induction we prove that $V_{k}(\Delta,q)$ is non-decreasing function for any $k$ with respect to $\Delta$ and any $q \in \left\{1,...,B-1\right\}$. By taking the limits on both side of \eqref{klimit} we prove that \eqref{lemma1_part1} holds in the case $0<q<B$.

\textbf{\emph{Case 3}}. $q=B$,

In this case, according to transition probability \eqref{transition_case1_v2} and \eqref{transition_case2_v2}, we have the state-value function $Q_k(\Delta, q,0)$ and $Q_k(\Delta, q,1)$ as follows:
\begin{align}
    Q_k(\Delta, q,0)=\Delta &+\lambda V_k(\Delta+1,q) \nonumber\\
    &+(1- \lambda)V_k(\Delta+1,q),
\end{align}
and,
\begin{align}
    Q_k(\Delta, q,1)=\Delta+\omega C_r &+p\lambda V_k(\Delta+1,q) \nonumber\\
    &+p(1- \lambda)V_k(\Delta+1,q-1)\nonumber\\
    &+(1-p)\lambda V_k(1,q)\nonumber\\
    &+(1-p)(1-\lambda) V_k(1,q-1).
\end{align}
Due to $V_{k}(\Delta,q)$ is assumed to be non-decreasing function with respect to $\Delta$ for any fixed $q$, it is obviously that both $Q_k(\Delta, q,0)$ and $Q_k(\Delta, q,1)$ are non-decreasing with respect to $\Delta$. Therefore,for any $\Delta_1\le\Delta_2$ we have:
\begin{align}
\label{V_K+1_DELTA1_lemma1_case3}
    V_{k+1}(\Delta_1,q) &= \min_{a \in \mathcal{A}}\left\{ Q_k(\Delta_1,q,a)\right\} \nonumber\\
    &=\min\left\{ Q_k(\Delta_1,q,0),Q_k(\Delta_1,q,1)\right\} \nonumber\\
    &\le \min\left\{ Q_k(\Delta_2,q,0),Q_k(\Delta_2,q,1)\right\} \nonumber\\
    &=V_{k+1}(\Delta_2,q).
\end{align}
As a result, with the induction we prove that $V_{k}(\Delta,q)$ is non-decreasing function for any $k$ with respect to $\Delta$ and $q=B$. By taking the limits on both side of \eqref{klimit} we prove that \eqref{lemma1_part1} holds in the case $q=B$.

To sum up, \eqref{lemma1_part1} holds and we complete the proof of the first part in Lemma\ref{lemma1}.

According to the exact same mathematical induction, we can also verify that the formula \eqref{lemma1_part2} holds. Due to limited space, the specific certification steps are omitted here. Thus we complete the proof of Lemma \ref{lemma1}.

\subsection{Proof of Lemma \ref{lemma2}}
\label{app_proof_lemma_creasement}

First, let's prove \eqref{lemm2_formula1}. By the \eqref{lemma1_part1} of Lemma \ref{lemma1}, assuming $\Delta_1 \le \Delta_2$ and $q \in \left\{1,...,B-1\right\}$, it is easy to yield
    \begin{align}
        Q(\Delta_2,q,0)&-Q(\Delta_1,q,0) =\Delta_2 - \Delta_1 \nonumber\\
        &+\lambda[V(\Delta_2+1,q+1)-V(\Delta_1+1,q+1)]\nonumber\\
        &+(1-\lambda)[V(\Delta_2+1,q)-V(\Delta_1+1,q)]\nonumber\\
        \ge &\Delta_2 - \Delta_1,
    \end{align}
    and,
    \begin{align}
        Q(\Delta_2,&q,1)-Q(\Delta_1,q,1) =\Delta_2 - \Delta_1 \nonumber\\
        &+p\lambda[V(\Delta_2+1,q)-V(\Delta_1+1,q)]\nonumber\\
        &+p(1-\lambda)[V(\Delta_2+1,q-1)-V(\Delta_1+1,q-1)]\nonumber\\
        &+(1-p)\lambda[V(1,q)-V(1,q)]\nonumber\\        
        &+(1-p)(1-\lambda)[V(1,q-1)-V(1,q-1)]\nonumber\\
        \ge &\Delta_2 - \Delta_1.
    \end{align}
Due to $V(\textbf{x})= \mathop{\min}\limits_{a \in \mathcal{A}} Q(\textbf{x},a)$, we prove that formula \eqref{lemm2_formula1} holds for all $q \in \left\{1,...,B-1\right\}$. Through the same proof process, it can also be verified that \eqref{lemm2_formula1} is also valid when $q=0$ and $q=B$. Therefore, we have proved $V(\Delta_2,q)-V(\Delta_1,q)\ge \Delta_2-\Delta_1$ holds for any $\Delta_1 \le \Delta_2$ and fixed $q \in \mathcal{B}$.

Second, we will tackle formula \eqref{lemm2_formula2}. The following proof needs to apply VIA and mathematical induction. For the convenience of explanation, an equivalent transformation is made to formula \eqref{lemm2_formula2} as follows:
    \begin{equation}
    \label{trans_lemma2_formula2}
        V(\Delta+1,q+1)+p V(\Delta,q)\ge V(\Delta,q+1)+p V(\Delta+1,q),
    \end{equation}
    
for state $\textbf{x}$, we have 
\begin{align}
V(\textbf{x})&= \mathop{\min}\limits_{a \in \mathcal{A}} Q(\textbf{x},a)\nonumber\\
&=\min\left\{ Q_k(\textbf{x},0),Q_k(\textbf{x},1)\right\}.
\end{align}

So every value function in \eqref{trans_lemma2_formula2} has two possible values. In order to prove formula \eqref{trans_lemma2_formula2}, theoretically we need to discuss $2^4=16$ cases, which is obviously a bit too cumbersome. Here we use a little trick, that is, as long as we prove that for the $2^2=4$ possible combinations on the left side of the inequality sign, there exists a combination on the right side of the inequality sign to make "$\ge$" hold, then we can prove formula \eqref{trans_lemma2_formula2}. Next, we make a mapping, using four numbers to sequentially represent the action taken by the minimum state-action value function in formula \eqref{trans_lemma2_formula2}, that is, "1010" represents the following:
    \begin{align}
    \label{1010}
        Q(\Delta+1,q+1,1)+pQ(\Delta,q,0)\ge \nonumber\\ Q(\Delta,q+1,1)+pQ(\Delta+1,q,0),
    \end{align}
So according to the previous trick, we only need to verify "0000", "1010", "0101", "1111" to prove formula \eqref{trans_lemma2_formula2}. Due to limited space, we only show the verification process of "1010" in the following proof. The other three cases can also be proved by the same steps.

Now we start to apply VIA. Assuming that $V_0(\textbf{x})=0$ for any states $\textbf{x}$, we have:
\begin{align}
    &Q_0(\Delta+1,q+1,1)+p Q_0(\Delta,q,0)\nonumber\\
    &-[Q_0(\Delta,q,1)+p Q_0(\Delta+1,q,0)]\nonumber\\
    =&\Delta+1+p(\Delta+\omega(1-u(q))C_r)\\
    &-[\Delta+p(\Delta+\omega(1-u(q))C_r)]\nonumber\\
    =&1 \ge 0.
\end{align}
Then for $Q_0(\textbf{x})$ we can also verify the same property in the "0000", "0101", "1111" case by the similar calculation, which implies:
\begin{equation}
\label{V_1_trans_lemma2_formula2}
    V_1(\Delta+1,q+1)+p V_1(\Delta,q)\ge V_1(\Delta,q+1)+p V_1(\Delta+1,q),
\end{equation}
for any $q \in \left\{0,1,...,B-1\right\}$ and $\Delta \in \mathbb Z^+ $. By induction, assuming that for any $q \in \left\{0,1,...,B-1\right\}$ and $\Delta \in \mathbb Z^+ $, we have:
\begin{equation}
\label{V_k_trans_lemma2_formula2}
    V_k(\Delta+1,q+1)+p V_k(\Delta,q)\ge V_k(\Delta,q+1)+p V_k(\Delta+1,q).
\end{equation}
What we need to do is to verify that formula \eqref{trans_lemma2_formula2} still holds in the next value iteration. Again, we take a look at the "1010" case. For $\Delta \in \mathbb Z^+ $ and $q \in \left\{0,1,...,B-1\right\}$, we have:
    \begin{align}
    \label{Qk11010}
        &Q_k(\Delta+1,q+1,1)+pQ_k(\Delta,q,0)\nonumber\\
        &-[Q_k(\Delta,q+1,1)+pQ_k(\Delta+1,q,0)]\nonumber\\
        =&\Delta+1+p\lambda V_k(\Delta+2,q+1)+p(1- \lambda)V_k(\Delta+2,q)\nonumber\\
    &+(1-p)\lambda V_k(1,q+1)+(1-p)(1-\lambda) V_k(1,q)\nonumber\\
    &+p[\Delta+\omega C_r+\lambda V_k(\Delta+1,q+1)+(1- \lambda)V_k(\Delta+1,q)]\nonumber\\
    &-\Delta-p\lambda V_k(\Delta+1,q+1)-p(1- \lambda)V_k(\Delta+1,q)\nonumber\\
    &-(1-p)\lambda V_k(1,q+1)-(1-p)(1-\lambda) V_k(1,q)\nonumber\\
    &-p[\Delta+1+\omega C_r+\lambda V_k(\Delta+2,q+1)-(1- \lambda)V_k(\Delta+2,q)]\nonumber\\
    =&1-p\ge 0.
    \end{align}

Therefore, by the similar step, we can verify the other three cases and get the following formula 
\begin{equation}
\label{V_k+1_trans_lemma2_formula2}
    V_{k+1}(\Delta+1,q+1)+p V_{k+1}(\Delta,q)\ge V_{k+1}(\Delta,q+1)+p V_{k+1}(\Delta+1,q)
\end{equation}
holds for any $\Delta \in \mathbb Z^+ $ and $q \in \left\{0,1,...,B-1\right\}$. By induction we confirm that for any $k$, the formula  \eqref{V_k_trans_lemma2_formula2} holds. Take the limits of $k$ on both side then we are able to prove that \eqref{trans_lemma2_formula2} holds, which is equivalent to \eqref{lemm2_formula2} holds. Hence, we complete the whole proof.

\end{document}